\documentstyle[12pt]{article}

\input feynman

\advance\textheight by \topskip
\newcommand{\be}{\begin{equation}}
\newcommand{\ee}{\end{equation}}
\newcommand{\br}{\begin{eqnarray}}
\newcommand{\bea}{\begin{eqnarray}}
\newcommand{\beanon}{\begin{eqnarray*}}
\newcommand{\er}{\end{eqnarray}}
\newcommand{\eea}{\end{eqnarray}}
\newcommand{\eeanon}{\end{eqnarray*}}
\newcommand{\ba}{\begin{array}}
\newcommand{\ea}{\end{array}}
\newcommand{\bi}{\begin{itemize}}
\newcommand{\ei}{\end{itemize}}
\newcommand{\bn}{\begin{enumerate}}
\newcommand{\en}{\end{enumerate}}
\newcommand{\bc}{\begin{center}}
\newcommand{\ec}{\end{center}}

\newcommand{\dotp}{\!\cdot\!}

\newcommand{\ar}{\rightarrow}

\newcommand{\Dir}{\kern -6.4pt\Big{/}}
\newcommand{\Dirin}{\kern -10.4pt\Big{/}\kern 4.4pt}
\newcommand{\DDir}{\kern -7.6pt\Big{/}}
\newcommand{\DGir}{\kern -6.0pt\Big{/}}
\begin{document}
\tolerance=100000
\thispagestyle{empty}
\setcounter{page}{0}

\begin{flushright}
{\large DFTT 78/93}\\
{\rm March 1994\hspace*{.5 truecm}}\\
\end{flushright}

\vspace*{\fill}

\begin{center}
{\Large \bf $e^+e^-\rightarrow b\bar bW^+W^-$ at the
Next Linear Collider, \mbox{ top--pairs} and Higgs production.
\footnote{Work supported in part by Ministero
dell' Universit\`a e della Ricerca Scientifica.\\[4. mm]
Mail address: V. Pietro Giuria 1, 10125 Torino, Italy;\\
e-mail: ballestrero,maina,moretti@to.infn.it.}}\\[2.cm]
{\large Alessandro Ballestrero, Ezio Maina, Stefano Moretti}\\[.3 cm]
{\it Dipartimento di Fisica Teorica, Universit\`a di Torino, Italy}\\
{\it and INFN, Sezione di Torino, Italy.}\\[1cm]
\end{center}

\vspace*{\fill}

\begin{abstract}
{\normalsize
\noindent
The complete matrix element for $e^+e^-\ar W^+W^-\bar b b$
is computed at tree level and applied to: {\bf a)} $\bar t t$ production and
decay ; {\bf b)} $ZH$ production followed by $Z\ar \bar bb$ and $H\ar WW$.
In both cases we include finite width effects and all irreducible
backgrounds.
}
\end{abstract}

\vspace*{\fill}
\newpage

\subsection*{Introduction}
The case for the construction of the Next--Linear--Collider (NLC), an
$e^+e^-$ accelerator with a total energy in the range between 300 GeV and
500 GeV, is quite convincing \cite{ee500,JLC}. There are
at least two fundamental topics, within the Standard Model (SM),
which can be explored at such a machine. First, we have learned, from
the comparison of the high precision measurements performed at the
$Z^0$ peak with the theoretical predictions
that the mass of the $top$ must be
large, confirming the results of direct searches at the Tevatron, but that it
is extremely difficult to reconcile a $t$--quark heavier than about
200 GeV with LEP data. Therefore we expect the $top$ to be discovered in the
not--too--distant future at Fermilab. With a $\bar t t$ threshold below
400 GeV, NLC would be in an ideal position to
study the physics of $top$, measuring its parameters with much greater
precision than it would be possible at hadron colliders.
Second, NLC would provide the possibility of studying the interactions of
longitudinally polarized $W$'s, and the delicate cancellations which are
related to the gauge structure
of the theory and which are essential to preserve unitarity. The magnitude of
these cancellations increases sharply with the center--of--mass energy and
will only be observed with difficulty at LEP II, where $W$ pairs will be
produced almost at rest.
Furthermore, if the Higgs boson is not too heavy, it might be possible
to explore in detail its couplings and parameters.\par
If the SM turned out to be only a low energy approximation to a larger
theory, NLC would be at the forefront of the exploration of
the new physics. For instance, in the Minimal Supersymmetric
Standard Model there are five Higgs states instead of the single Higgs particle
predicted in the SM. Some of them are necessarily of relatively low mass
and would certainly be detected at NLC.\par
In this letter we examine the production of $W^+W^-\bar b b$ in $e^+e^-$
collisions, taking into account the full set of
diagrams, shown in fig.1, which describe this reaction at tree level.\par
Our approach allows us to study the effects of the finite width of the $top$
and of the irreducible
background to $\bar t t$ production, including their interference.
It also includes correlations between the decays of the two $t$--quarks.
All previous studies of $\bar t t$ production have been performed
separating the production of a $\bar t t$ pair from the independent decay
of the two heavy quarks, in the spirit of the so--called
narrow width approximation (NWA), even though some finite width effects
have been included in several instances.
Electroweak and QCD higher order
corrections to the two subprocesses have been computed
\cite{widthtop,qcdcorr,ewcorr}.
$\Gamma_t$ grows as the third power of the $top$ mass and, given the present
limits on $m_t$, the lifetime is of the order of
$10^{-23}$ $sec$ and the $top$ decays before hadronizing.
One expects finite width effects at the percent level and
it is important, therefore, to fully include
them in the theoretical analysis.\par
Several strategies, depending on $E = \sqrt{s} - 2 m_t$,
have been proposed \cite{ee500exp} in order to measure the mass
of the $top$ and possibly its other parameters as well.
Most of the theoretical activity has focused on
the threshold region which
allows very precise measurements, provided a suitably
small beam energy spread can be achieved.
The excitation curve and the momentum distributions are
dominated by the QCD analogue of the Coulomb attraction
and can be reliably computed
\cite{ttthreshold1,ttthreshold2,ttthreshold3,ttthreshold4,ttthreshold5}.
The effects of the binding potential, however,
decrease rapidly above threshold \cite{ttthreshold3} and for $E \approx 2$ GeV
the two $t$--quarks behave as free particles.
Our calculation is clearly complementary to the methods
used in
\cite{ttthreshold1,ttthreshold2,ttthreshold3,ttthreshold4,ttthreshold5}.
It applies to all energies excluding a narrow window of a few GeV at threshold.
\par
A wealth of information, in addition to a precise measurement of the
$top$ mass can be extracted from the data and
it is important to have several measurements
performed with independent methods.
{}From the comparison of different observables one can
determine $\alpha_s$,  possibly measure
the lifetime of the $t$--quark and one can hope to gain information
on the coupling of the $top$ to the Higgs.
For instance,
the value of $m_t$ extracted from the $\bar tt$
excitation curve is strongly
correlated with the value of $\alpha_s$, and a high precision,
independent measurement of the mass would allow
a precise determination of the strong coupling constant
at $Q^2 =m_t^2$.
Such a measurement could be provided by an analysis of the
width of the momentum spectrum of the $W$'s produced in the
decay of the $top$. The expected uncertainty on $m_t$
is about 500 MeV \cite{stiegler,ee500exp}.
The direct reconstruction of $m_t$ from jets is difficult in
the threshold region because $\bar t t$ events are highly spherical
and the particles
from the two decays tend to mix. At higher energies this procedure
is simpler and one can hope to reconstruct the mass of the $top$
from the invariant mass distribution of combinations of three jets.
\par
The most important backgrounds to the $\bar t t$ signal
arise from multi--jet events, $Z^0\ar \bar q q + gluons$,
and from $W^+W^-$ and $Z^0Z^0$ pair production. The tagging of
$\bar t t$ events \cite{ee500exp} can be obtained exploiting their spherical
shape, the large number of jets, the presence of $b$--quark
jets signalled by isolated leptons or revealed in vertex
detectors. All these methods are ineffective against the irreducible
background given by non--resonant production of $W^+W^-\bar b b$
and a study of all mechanisms leading to this final
state is needed.
\par
We have concentrated our analysis on $\bar t t$ production, neglecting the
interesting possibility of single $top$ production
\cite{singletop}, which would be relevant if
in the first period of operation the energy of NLC were smaller than $2m_t$.
\par
The full matrix element for $e^+e^-\ar WW\bar b b$ includes the
production of a SM Higgs in $e^+e^-\ar ZH$ followed by $Z\ar \bar b b$
and $H\ar WW$.
At center of mass energy of about 300 GeV the Higgs is produced in association
with a $Z^0$ much more efficiently than in $WW$
or $\gamma\gamma$ fusion. As the energy is increased the $WW$--fusion mechanism
becomes more relevant and at 500 GeV the two processes have comparable cross
sections over most of the Higgs mass range which can be probed at NLC.
It is important to detect the Higgs produced in both kind
of reactions in order to measure its coupling to both $W$'s and $Z^0$'s.
For $M_H$ above threshold the decay mode $H\ar WW$ is much larger than the
$\bar b b$ channel which dominates at smaller $M_H$.
It is to be stressed that the $\bar b b$ decay might well be one of best
ways to detect the $Z^0$, thanks to the expected performance of vertex
detectors. This mode has a branching ratio
about five times larger than the branching ratios to $\mu^+\mu^-$
or $e^+e^-$ and it is equally free from backgrounds coming from $W$ decays.
The possibility of studying Higgs production using the even larger
invisible decay mode $Z\ar \bar \nu \nu$ has been recently examined in
ref. \cite{wisc}.
The cross section for $ZH\ar WW \bar bb$ production is much smaller
than the cross section for $\bar t t \ar WW \bar bb$ production and therefore
it is essential in order to asses the observability of the former
to have a complete description of the latter and of the interference between
the two.\par

\subsection*{Calculation}
The matrix element for $e^+e^-\ar W^+W^-\bar b b$
has been computed using the method presented in \cite{method}
and, as a check, in the formalism of \cite{hz}. In the latter case we
have used the unitary, Landau and Feynman gauges.
We have checked that our matrix element squared
does not depend on the choice of the auxiliary vectors
$k_0$ and $k_1$.
We found that the program using the formalism \cite{method} was
about four times faster than the program using the formulae of
ref.\cite{hz}.\par
In the methods of \cite{method,hz} the analytical expressions of the
diagrams is of little
use, since they amount to a textbook application of the Feynman rules,
and therefore we do not give them.\par
We have verified that in the NWA our results
reduce smoothly to the cross section obtained from $e^+e^-\ar \bar t t$.
For this we have taken into account only diagram number 2 in fig.1,
writing the heavy quark propagator as
\be
\frac{ p\Dir  + m}{p^2-m^2+im\Gamma}
\left( \frac{\Gamma}{\Gamma_{tot}}\right)^{\frac{1}{2}}.
\ee
When $\Gamma = \Gamma_{tot}$, the total width,
the standard expression is recovered.
The NWA can be obtained
taking numerically the limit $\Gamma \ar 0$. In practice
the cross section coincides with
the one computed as production times branching ratios
already for $\Gamma = 1\times 10^{-3}$.
All our results in the NWA have been obtained with
$\Gamma = 1\times 10^{-5}$.\par
For the width of the $t$--quark and of the Higgs boson
we have used their tree level expressions.
This seems to us the choice most coherent with our treatment
of the whole process, particularly in relation to the
fact that $top$ width is actually smaller when QCD corrections
are included than at tree level.
In addition this is necessary in order to recover the
$e^+e^-\ar \bar t t$ cross section in the NWA, since we do not
include QCD corrections to the decay process.\par
In order to better control the interplay of the various peaks
in phase space, we have integrated separately in the
Higgs region, $\mid M_{WW} - M_H\mid < 4\times\Gamma_H$
and $\mid M_{\bar b b} - M_Z\mid < 4\times\Gamma_Z$,
and in its complement.
We have used different integration variables in the two cases,
grouping together the particles which in each region
are produced in resonant channels and mapping as appropriate the
narrow peaks which dominate the two contributions to the cross section.\par
We have used  $M_Z=91.1$ GeV, $\Gamma_Z=2.5$ GeV,
$\sin^2 (\theta_W)=.23$, $M_W=M_Z\cos (\theta_W)$, $\Gamma_W=2.2$ GeV,
$m_b=5.$ GeV, $\alpha_{em}= 1/128$ in the numerical part of our work.\par
The $b$--quark has been treated as a massive stable particle, while
the widths of all virtual $t$--quark, $W$, $Z$ and Higgs bosons
have been included in all our calculations.
We have not included the effects of the width of the final state $W$'s
which is known to be non--negligible \cite{W_width}. It would be relatively
easy to substitute in our formulae the polarization vectors of the $W$'s
with their decay amplitude including the complete propagator.
We don't know, at this time, whether the complications of six additional
integration variables,
with two more sharp peaks in phase space can be dealt with using a reasonable
amount of CPU time. It appears extremely difficult to compute
the full amplitude for $\bar b b jjjj$ or $\bar b b \nu\ell jj$ production
with all combinations of quark and gluon jets.
\par

\subsection*{The matrix element}

We give in the following a description of how the calculation of the matrix
element has been performed with the method
of ref.~\cite{method}. To this end we will consider only diagrams 1-19
not containing Higgs propagators. The remaining ones can be easily dealt with
with similar
techniques. We use the notation and symbols introduced in
ref.~\cite{method}  and we refer to it for the basics of the method.
We denote the four momenta of the positron and the electron as $p_1$ and $p_2$,
respectively, those of  $b$ and $\bar b$ as $q_1$ and $q_2$
and those of $W^+$ and $W^-$ as $r_1$ and $r_2$. The index $k$ will label
the polarization states of electron and positron, $l$ and $m$ those of
$b$ and $\bar b$,  $i$ and $j$  those of $W^+$ and $W^-$.
\par
We have used the
polarizations of ref.~\cite{hz} for the external vector particles,
$\eta^+_{i\mu}$ for $W^+$ and $\eta^-_{j\mu}$ for $W^-$.
We have computed, with standard Feynman
diagrams techniques, the vectors $\eta^{WWV}_{ij\mu}$ ($V=\gamma ,Z$),
corresponding to the insertion in fermion lines of a $\gamma$ and a $Z$
connected through a triple vector coupling to the external $W$'s (see diagrams
9, 10, 11 12).
\par
To obtain all remaining vectors $\eta_{\mu}$ corresponding to the insertion in
the massive fermion line of $\gamma$, $Z$ or $W$ propagators connected to the
electron spinor line, one has to start  by computing
$A_{kk}(p_1;\mu,c^{eV};p_2)$ ($V=\gamma ,Z$).
{}From these, one gets
\be
\eta_{k\mu}^{\gamma}=\frac{A_{kk}(p_1;\mu,c^{e\gamma};p_2)}{s} \hskip 2truecm
\eta_{k\mu}^{Z}=\frac{A_{kk}(p_1;\mu,c^{eZ};p_2)}{s-M^2_Z+i\Gamma_Z M_Z}
\ee
which are the vectors to be used in diagrams 1,2,3,9,10. Multiplying these
for the suitable propagators and vertices and contracting with $\eta^+_{i\mu}$
and/or $\eta^-_{j\mu}$, according again to standard Feynman rules, we
have computed
$\eta_{kij\mu}^{13V}$, the $\eta$ corresponding to the $V$ ($V=\gamma ,Z$)
insertion on the quark line relative to diagram 13, and the analogous
quantities $\eta_{kij\mu}^{14V}$, $\eta_{kij\mu}^{15V}$, $\eta_{ki\mu}^{16W}$,
$\eta_{kj\mu}^{17W}$.
\par
All $\tau$ matrices corresponding to all other insertions
in the upper, central or lower position of the massless spinor line have
then been computed. Besides
the insertions of the vectors $\eta^+_{i\mu}$, $\eta^-_{j\mu}$,
$\eta^{WWV}_{ij\mu}$, one has also those corresponding to $\gamma$, $Z$ or
$W$ propagators. For these, one $\tau$ matrix for every Lorentz index has to
be evaluated. For instance, the matrix of the lower $\gamma$ insertion of
diagram~4 is $\tau (p_1;\mu,c^{e\gamma};p_2-r_1-r_2)$. It has to be noticed
that, as
already pointed out in ref.~\cite{method}, only the $A$ and $C$ components of
the lower
insertions and the $A$'s and $B$'s of the upper ones are needed. This is
because, considering the electrons  massless, one wants only the $A$ component
of the matrix corresponding to the whole electron line.
To get for example the $A$ of diagram~7 one computes:
\bea
A^{7W}_\mu &=& A(p_1;\mu,c^{eW\nu};p_2-r_2;\eta^-_{j\mu};c^{eW\nu};p_2)
\nonumber\\
&= & A(p_1;\mu,c^{eW\nu};p_2-r_2)\; A(p_2-r_2;\eta^-_{j\mu};c^{eW\nu};p_2)\\
&  &+C(p_1;\mu,c^{eW\nu};p_2-r_2)\; B(p_2-r_2;\eta^-_{j\mu};c^{eW\nu};p_2).
\nonumber
\eea
All other global $A$'s for the electron line have been determined in
similar manner according to the rules of ref.\cite{method}.
These quantities must then be divided for
a factor $p^2\,p\dotp k_0$ for every piece of massless fermion propagator
carrying momentum $p$, and they must be multiplied for the appropriate
propagators and vertices and eventually combined with $\eta^+_{i\mu}$
or $\eta^-_{j\mu}$ (see diagrams 18,19), in order to obtain all $\eta$'s
for all insertions in the quark spinor line.
To compute directly $A^{4V}_\mu/\left[(p_2-r_2)\dotp k_0\;(p_2-r_2)^2\right]
+A^{11V}_\mu$, we found convenient
to introduce
\bea
\tau^\prime&=&\frac{\tau
(p_2-r_1-r_2;\eta^+_i,c^{eW\nu};p_2-r_2;\eta^-_j,c^{eW\nu};p_2)}
{(p_2-r_2)\dotp k_0\;(p_2-r_2)^2}\\
& & +\tau (p_2-r_1-r_2;\eta^{WW\gamma}_{ij},c^{e\gamma};p_2)
+\tau (p_2-r_1-r_2;\eta^{WWZ}_{ij},c^{eZ};p_2)\nonumber
\eea
and then compose $\tau^\prime$ with $\tau(p_1;\mu,c^{eV};p_2-r_1-r_2)$.
Analogous considerations apply to $A^{6V}_\mu$ and $A^{12V}_\mu$.
\par
Having obtained all $\eta$'s, we have summed together the ones which have the
same
polarization indexes and enter the quark line with the same propagator:
\beanon
\eta_{kij\mu}^{T\,\gamma}&=&
(\eta_{kij\mu}^{4\,\gamma}+\eta_{kij\mu}^{11\,\gamma})
+(\eta_{kij\mu}^{6\,\gamma}+\eta_{kij\mu}^{12\,\gamma})+
\eta_{kij\mu}^{13\,\gamma}+\eta_{kij\mu}^{14\,\gamma}
+\eta_{kij\mu}^{15\,\gamma}+\eta_{kij\mu}^{18\,\gamma}
+\eta_{kij\mu}^{19\,\gamma}\\
\eta_{kij\mu}^{T\,Z}&=&
(\eta_{kij\mu}^{4\,Z}+\eta_{kij\mu}^{11\,Z})
+(\eta_{kij\mu}^{6\,Z}+\eta_{kij\mu}^{12\,Z})+
\eta_{kij\mu}^{5\,Z}+
\eta_{kij\mu}^{13\,Z}+\eta_{kij\mu}^{14\,Z}
+\eta_{kij\mu}^{15\,Z}+\eta_{kij\mu}^{18\,Z}+\eta_{kij\mu}^{19\,Z}\\
\eta_{kj\mu}^{T\, W^-}&=&
\eta_{kj\mu}^{7\, W^-}+\eta_{kj\mu}^{17\, W^-}\hskip 1.5truecm
\eta_{ki\mu}^{T\, W^+}=
\eta_{ki\mu}^{8\, W^+}+\eta_{ki\mu}^{16\, W^+}.
\eeanon
All $\tau$ matrices
corresponding to a single insertion of $\eta_{k\mu}^\gamma$,
$\eta_{k\mu}^Z$, $\eta^+_{i\mu}$, $\eta^-_{j\mu}$,
$\eta^{WW\gamma}_{ij\mu}$, $\eta^{WWZ}_{ij\mu}$ ,
$\eta_{kij\mu}^{T\,\gamma}$, $\eta_{kij\mu}^{T\,Z}$,
$\eta_{kj\mu}^{T\, W^-}$, $\eta_{ki\mu}^{T\, W^+}$ in the various allowed
positions of the quark line, have then been evaluated.
\par
The matrices $\tau(q_1;\eta_{kij}^{T\, V},c^{bV};q_2)$ lead immediately
to the amplitudes relative to diagrams 4, 5, 6, 11,12, 13, 14, 15,
18, 19, making use of the usual relation:
\be\label{amp}
 Amp= A +m_b B-m_b C-m_b^2 D.
\ee
Composing together $\tau_1=\tau(q_1;\eta_{ki}^{T\, W^+},c^{btW};-r_2-q_2)$
with $\tau_2=\tau(-r_2-q_2;\eta_{j}^-,c^{btW};q_2)$ one gets (cf. eqs.~(31,32)
ref.\cite{method}):
\be
\tau(q_1;\eta_{ki}^{T\, W^+},c^{btW};-r_2-q_2;\eta_{j}^-,c^{btW};q_2)=
\tau_1\bullet\tau_2
\ee
from which the diagrams $8+16$ are obtained. Analogously from
$\tau(q_1;\eta_{i}^+,c^{btW};q_1-r_1)$
and $\tau(q_1-r_1;\eta_{kj}^{T\, W^-},c^{btW};q_2)$ one gets diagrams $7+17$.
\par\noindent
To evaluate  diagrams $1+9$ one first computes $\tau(q_1;\eta_i^+,
c^{btW};q_1-r_1;\eta_{j}^-,c^{btW};q_1-r_1-r_2)$. From this one derives
the matrix
\bea
\tau^{WW}_{up}&=&\frac{\tau(q_1;\eta_i^+,c^{btW};q_1-r_1;
\eta_{j}^-,c^{btW};q_1-r_1-r_2)}{(q_1-r_1)\dotp k_0\;
\left[(q_1-r_1)^2-m_t^2+ i \Gamma_t m_t\right]}\\
& &+\tau (q_1;\eta^{WW\gamma}_{ij},c^{e\gamma};q_1-r_1-r_2)
+\tau (q_1;\eta^{WWZ}_{ij},c^{eZ};q_1-r_1-r_2)\nonumber
\eea
and subsequently
\be
\tau^{(1+9)}=\tau^{WW}_{up}\bullet \left(\tau(q_1-r_1-r_2;\eta^\gamma _k,
c^{e\gamma};q_2)+\tau(q_1-r_1-r_2;\eta^Z _k,c^{eZ};q_2)\right).
\ee
The $A$, $B$, $C$, $D$ elements of $\tau^{(1+9)}$, divided by
\[
(q_1-r_1-r_2)\dotp k_0\;\left[(q_1-r_1-r_2)^2-m_t^2+ i \Gamma_t m_t\right]
\;\sqrt{p_1\dotp k_0\,
p_2\dotp k_0\,q_1\dotp k_0\,q_2\dotp k_0}
\]
 give the amplitudes
relative to diagrams $1+9$ using eq. (\ref{amp}).
Finally the diagrams $3+10$ are evaluated with a procedure quite analogous
to that just described for diagrams $1+9$.
\par

\subsection*{Results and discussion}
Our results are given in fig.2 through fig.8 and in table I
and II.\par
In fig.2 we present the momentum distribution of the $W$'s at $\sqrt{s}=300$
GeV for $m_t =$ 145, 149, 149.5 GeV. In the upper part we show the results
obtained from the full amplitude and in the lower half the results obtained
in the NWA. It has been proposed \cite{ee500exp},
in order to enhance the $\bar t t$ signal, in events in which
one of the $W$'s decays leptonically,
to cut on the invariant mass $M_{had}$
of the hadronic
$t\bar b\ar jets$ system. The difference between the dashed and the continuous
lines is the effect of imposing $M_{had} > 180$ GeV for $m_t = 145$ GeV.
At larger values of $m_t$ the extra cut does not significantly affect the
results. Please notice that the distributions in fig.2 are
normalized to the total cross section, they do not
include the leptonic branching ratio of the $W$.
The sharp edges of the distribution obtained in the NWA are considerably
softened in the full calculation, particularly when the available phase space
is very small, and long tails of increasing size are formed.
The total cross section, given in table I decreases by
5$\div$8\% moving from the NWA (first column) to the full amplitude
(third column) while the peak height is
more strongly affected, decreasing by 20\% (30\%) for $m_t=$ 149 (149.5) GeV.
The comparison between the second and the third column shows that the
irreducible
background, with the exclusion of the Higgs contribution,
constitutes 1$\div$2\% of the total cross section.
\par
The measurement of the $top$ width appears much more difficult
then the measurement of its mass \cite{ee500exp}. Our calculation allows us to
study the sensitivity of different observables to $\Gamma_t$.
In fig.3 we show d$\sigma$/d$p_W$ with $\sqrt{s}=300$
GeV and $m_t = 149$ GeV for three values of the $top$ width .
The SM prediction (continuous line)
is compared with the predictions for a width 25\% (dashed) and 50\% (dotted)
larger.
Since we are assuming to detect the
decay $t\ar bW$, it is inconsistent to consider a width smaller than required
by this very decay channel. It is however conceivable
that additional channels exist, resulting in a width larger than predicted by
the SM. Obviously the cross section, for larger widths, decreases.
In the NWA one expects
\be
\frac{\sigma(\Gamma_1)}{\sigma(\Gamma_2)} = \frac{\Gamma_2^2}{\Gamma_1^2}.
\ee
Actually there is an additional reduction for larger widths
due to the limited
phase space which is available, since the high--invariant--mass tail
of the $top$ decay
distribution can fall outside the kinematical bounds.
This effect decreases the total cross section computed in the NWA
by about 2$\div$3\% and the peak height by approximately 4$\div$8\%, the larger
width
obviously giving the larger correction.\par
Before drawing any firm conclusion from these results they should be folded
with a realistic simulation of the expected performance of NLC detectors,
but they indicate that the finite width of the $top$ and the irreducible
background have a significant impact on the measurement of the parameters
of the $top$ from the $W$
momentum distribution.\par
At larger energies it becomes feasible to reconstruct the $top$ from
the jets in which it decays. When one of the $W$'s decays leptonically
four jets are expected in the final state, two from the $b$'s and two
from the
decay of the second $W$. If both $W$'s decay hadronically one expects six jets.
If the pair or pairs of jets from $W$ decay can be identified,
using their total
invariant mass, one can construct, for each $W$, two $t$--candidates
combining it with the two $b$--jets.
One of the combinations is expected to
peak at the $t$--quark mass while the other
builds up a wide distribution without any peak.
The value of the mass of the $top$ can be extracted from a fit of the observed
mass distribution, taking into account all experimental
and accelerator--related uncertainties. In fig.4 we present the mass
distribution for the ``right'' ($W^+b$ and $W^-\bar b$) and ``wrong''
($W^-b$ and $W^+\bar b$) three--jet combinations
for two representative values of the $top$ mass, $m_t = 130,\, 180$ GeV
at $\sqrt{s} = 500$ GeV.
In table II we give, in the third column, the
total cross sections for $m_t = 130,\, 150,\, 180,\, 200$ GeV
and compare them with the predictions of the NWA and with the
results obtained from diagram number 2 using the tree level $top$ width,
shown in the first and second column respectively.
For this set of results all diagrams involving the Higgs boson have been
set to zero. The cross section obtained from the larger set of diagrams
exceeds the results from the single diagram by about $7\div 9$\% at
$m_t = 130$ GeV and by $2\div 4$\% at $m_t = 200$ GeV.
This provides an estimate of the irreducible bacground to $\bar t t$
production in this energy range.\par
A Higgs boson of mass between $2 M_W$ and approximately
$\sqrt{s} - 100 $ GeV would certainly show up as a peak in the
$WW$ mass spectrum. The accompanying $Z^0$ in the $HZ$ final state
can be searched for
in all possible decay channels. From our point of view it is
particularly interesting the appearance of
a bump in the $\bar b b$ spectrum.
In fig.5 and 6 we show the mass distribution of the $\bar b b$ pairs
and of the $WW$ pairs at $\sqrt{s} = 300$ GeV for $m_t = 145$ GeV and
$m_t = 149$ GeV. The corresponding distributions at $\sqrt{s}= 500$ GeV
for $m_t = 130$ GeV and $m_t = 180$ GeV are given in fig.7 and 8.
In fig.5 and 6 the lower curves in the two sets on the left
(continuous and dash--dotted lines)
have been obtained with the restriction $M_{had}> 180$ GeV,
while in fig.7 and 8 they
have been obtained with the
requirement that the energies of two $Wb$ pairs
fall in the range $\mid E_{Wb}- E_{beam}\mid < .05\times E_{beam}$.
In the central part of fig.5 and 6 we present the result assuming a
Higgs mass $M_H = 170$ GeV, slightly above the threshold for the Higgs decay
to two real $W$'s. In fig.7 and 8 the Higgs mass is assumed to be 200 GeV.
In the inserts we compare these results,
in small regions around the $Z^0$ and the Higgs
resonances, with the histograms which result if all diagrams
involving the Higgs are set to zero.
The meaning of the different lines is described in the figure captions.
The $Z^0$ peak in the $\bar b b$ mass distribution and the Higgs peak
in the $WW$ mass distributions are perfectly visible with our mass resolution
of 1 GeV. They would still be observable with a resolution two or three
times worse. \par
A more quantitative estimate of the relevance of the background
from $\bar t t$ production to the detection of the Higgs in the
$WW\bar b b$ channel can be obtained comparing the cross section
$\sigma_T$ obtained from the full amplitude with the
cross section $\sigma_R$ obtained
from the resonant diagram, number 26 in fig.1, in the
phase space window defined by
$\mid M_{\bar b b} - M_Z\mid < 10.5$ GeV and
$\mid M_{WW} - M_H\mid < 1.5$ GeV for $M_H = 170$ GeV
or $\mid M_{WW} - M_H\mid < 6.5$ GeV for $M_H = 200$ GeV.
The marked difference in the limits on $\mid M_{WW} - M_H\mid$
reflects the rapid increase of the Higgs width in the mass region
near the thresholds for the decay to real vector bosons.
At $\sqrt{s} = 300$ GeV, with $M_H = 170$ GeV we obtain
$\sigma_R = 12.9$ $fb$ and $\sigma_T = 22.7 \, (17.0)$ $fb$
for $m_t = 145 \, (149)$ GeV showing that the background
can be substantial.
On the contrary, at $\sqrt{s} = 500$ GeV, with $M_H = 200$ GeV
the background is very small. We have
$\sigma_R = 4.0$ $fb$ and $\sigma_T = 4.4$ $fb$
for both $m_t = 130$ GeV and $m_t = 180$ GeV.\par
The small peaks which can be seen in the inserts of fig.7, for
$M_{\bar b b} = M_{Z^0}$, even in the absence of the Higgs correspond
to $e^+e^-\ar WWZ\ar \bar b b WW$. The total cross section, both with
and without Higgs, are in rough agreement with the result given in
\cite{ee500WWZ} times the branching ratio $B(Z\ar \bar bb)$.
Notice however that the $\bar t t$ background is rather large.
At $\sqrt{s} = 300$ GeV,
if the Higgs is too heavy to be produced in association with the $Z^0$,
no peak at the $Z^0$ mass can be seen
in the $\bar b b$ spectrum.
The cross section for $e^+e^-\ar WWZ$ is too small at this energy.
\par
A few more comments are in order before concluding our discussion.
The $WW$ mass distribution presented in fig.8 has a number of
interesting features. It has an obvious dependence on the $top$ mass,
with a well defined triangular shape,
and it is rather broad. Therefore it will be rather insensitive to
the experimental uncertainties in measuring momenta. It might be used to
determine $m_t$ from the vast majority of events
in which both $W$'s decay hadronically.
Furthermore the $WW$ spectrum is very small for masses close to the
nominal center--of--mass energy. Therefore a cut on the $WW$ mass
could be quite efficient for the separation of the
$\bar t t$ signal from the production of $WW$ and $ZZ$ pairs.\par

\subsection*{Conclusions}
We have produced the complete matrix element for
$e^+e^-\ar \bar b b WW$ at tree level, using a new, more efficient method
for computing helicity amplitudes. We have discussed the production of
$\bar t t$ pairs at a high energy $e^+e^-$ collider avoiding the separation
of the production process from the subsequent decays.
Finite width effects, irreducible backgrounds and correlations between the
two $t$--quark decays are included in our treatment.
Our results are applicable to all energies excluding a narrow
window of a few GeV at threshold.\par
We have found that the momentum spectrum of the $W$'s, slightly above
threshold is strongly affected by the finite width of the $top$, confirming
earlier results based on the QCD potential.\par
In a number of cases we have studied the possibility of detecting the SM Higgs
produced in the Bjorken process $e^+e^- \ar HZ$ followed by $Z\ar \bar b b$
and $H\ar WW$. The background from $\bar t t$ production is not forbiddingly
large.\par

\vfill

\newpage

\subsection*{Figure Captions}
\begin{description}
\item[fig.1 ] Feynman diagrams contributing in the lowest order to
$e^+e^-\rightarrow b\bar b W^+W^-$.
Internal wavy--lines represent a $\gamma$, a $Z^0$ or a $W^\pm$,
as appropriate. Internal dashed--lines represent an Higgs boson $H^0$.
Diagrams for final $up$--fermions can be easily
obtained by exchanging $W^\pm$ bosons and trivially modifying
the $W^\pm$--bremsstrahlung topology in diagrams (7)--(8) and (16)--(17).

\item[fig.2 ] The momentum distribution of the $W$'s at $\sqrt{s} = 300$ GeV
for $m_t =$ 145 GeV (dashed and continuous), 149 GeV (dotted) and
149.5 GeV (dot--dashed). The curves in the upper part have been obtained
from the full amplitude, those in the lower part from diagram 2 only,
in the NWA. The continuous line gives the distribution if the additional
cut $M_{had} > 180$ GeV is imposed, where $M_{had}$ is the mass of the $tb$
system, in events in which one of the $W$'s decays leptonically.
For the larger values of $m_t$ the effect of the cut is irrelevant.

\item[fig.3 ] The momentum distribution of the $W$'s at $\sqrt{s} = 300$ GeV
for $m_t =$ 149 GeV for different choices of the $top$ width.
The continuous curve has been obtained using the
SM tree--level width $\Gamma_{SM}$. The dashed and dotted lines
have been obtained using $\Gamma = 1.25\times\Gamma_{SM}$ and
$\Gamma = 1.5\times\Gamma_{SM}$ respectively.

\item[fig.4] Mass distribution of the ``right''
($W^+b$ and $W^-\bar b$) and ``wrong''
($W^-b$ and $W^+\bar b$) three--jet combinations
at $\sqrt{s} = 500$ GeV for $m_t = 130,\, 180$ GeV.

\item[fig.5] The $\bar b b$ mass distribution
at $\sqrt{s} = 300$ GeV for $m_t = 145$ GeV and
$m_t = 149$ GeV.
The main figure describes the distribution assuming a
Higgs of mass $M_H = 170$ GeV. We study the effect, in events with
one $W$ decaying leptonically, of requiring an hadronic mass of at least
180 GeV
(continuous and dash--dotted lines) comparing with the results
obtained when this cut is
not enforced (dashed and dotted lines). In the inserts we show
the small region in phase space in which $M_{\bar b b} \approx M_{Z^0}$.
The histograms, from top to bottom, correspond to the four cases:
a) Higgs+no--cut
b) Higgs+cut
c) no--Higgs+no--cut
d) no--Higgs+cut
for $m_t = 145$ GeV (upper part) and
$m_t = 149$ GeV (lower part).

\item[fig.6] The $WW$ mass distribution
at $\sqrt{s} = 300$ GeV for $m_t = 145$ GeV and
$m_t = 149$ GeV.
The main figure assumes a
Higgs of mass $M_H = 170$ GeV. We study the effect, in events with
one $W$ decaying leptonically, of requiring an hadronic mass of at least
180 GeV
(continuous and dash--dotted lines) comparing with the results
obtained when this cut is
not enforced (dashed and dotted lines). In the inserts we show
the small region in phase space in which $M_{WW} \approx M_{H}$.
The histograms correspond to the four cases:
a) Higgs+no--cut (dashed)
b) Higgs+cut (continuous)
c) no--Higgs+no--cut (dotted)
d) no--Higgs+cut (dash--dotted)
for $m_t = 145$ GeV (upper part) and
$m_t = 149$ GeV (lower part).

\item[fig.7] The $\bar b b$ mass distributions
at $\sqrt{s} = 500$ GeV for $m_t = 130$ GeV and
$m_t = 180$ GeV.
The main figure describes the distribution assuming a
Higgs of mass $M_H = 200$ GeV. We study the effect, in events with
one $W$ decaying leptonically, of the additional cut
$\mid E_{Wb}- E_{beam}\mid < .05\times E_{beam}$
(continuous and dash--dotted lines) comparing with the results
obtained when this cut is
not enforced (dashed and dotted lines). In the inserts we show
the small region in phase space in which $M_{\bar b b} \approx M_{Z^0}$.
The histograms, from top to bottom, correspond to the four cases:
a) Higgs+no--cut
b) no--Higgs+no--cut
c) Higgs+cut
d) no--Higgs+cut
for $m_t = 130$ GeV (upper part) and
$m_t = 180$ GeV (lower part).
The requirement that the energy of the $Wb$ system to be close to half
the total energy washes out the Higgs peak almost completely.

\item[fig.8] The $WW$ mass distribution
at $\sqrt{s} = 500$ GeV for $m_t = 130$ GeV and
$m_t = 180$ GeV.
The main figure assumes a
Higgs of mass $M_H = 200$ GeV.  We study the effect, in events with
one W decaying leptonically, of the additional cut
\mbox{$\mid E_{Wb}- E_{beam}\mid < .05\times E_{beam}$}
(continuous and dash--dotted lines) comparing with the results
obtained when this cut is
not enforced (dashed and dotted lines). In the inserts we show
the small region in phase space in which $M_{WW} \approx M_{H}$.
The histograms correspond to the four cases:
a) Higgs+no--cut (dashed)
b) Higgs+cut (continuous)
c) no--Higgs+no--cut (dotted)
d) no--Higgs+cut (dash--dotted)
for $m_t = 130$ GeV (upper part) and
$m_t = 180$ GeV (lower part).

\end{description}

\vfill

\subsection*{Table Captions}
\begin{description}
\item[Table I]
Total cross section in $fb$ for
$e^+e^-\rightarrow b\bar bW^+W^-$ at $\sqrt{s} = 300$ GeV
with $m_t = 145,\, 149,\, 149.5$ GeV.
The three column refer, from left to right, to the
NWA, to the
results obtained from diagram number 2 using the tree level $top$ width
and to the full amplitude respectively.
No cut has been applied to enhance the $\bar t t$ signal.
For this set of results all diagrams involving the Higgs boson have been
set to zero.

\item[Table II]
Total cross section in $fb$ for
$e^+e^-\rightarrow b\bar bW^+W^-$ at $\sqrt{s} = 500$ GeV
with $m_t = 130,\, 150,\, 180,\, 200$ GeV.
The three column refer, from left to right, to the
NWA, to the
results obtained from diagram number 2 using the tree level $top$ width
and to the full amplitude respectively.
No cut has been applied to enhance the $\bar t t$ signal.
For this set of results all diagrams involving the Higgs boson have been
set to zero.

\end{description}

\vfill
\newpage

\pagestyle{empty}
\
\vskip2.0cm
$$\vbox{\tabskip=0pt \offinterlineskip
\halign to380pt{\strut#& \vrule#\tabskip=1.em plus2.0em& \hfil#&
\vrule#& \hfil#&
\vrule#& \hfil#&
\vrule#& \hfil#&
\vrule#\tabskip=0pt\cr \noalign{\hrule}
&  && && && && \cr
&  && &$t\bar t\hskip1.2cm$& &~$t\bar t \rightarrow \bar b bW^+W^-$& &$\bar b
bW^+W^-$~~~& \cr
&  && && && && \cr
\noalign{\hrule}
&  && && \omit && \omit && \cr
&  &$m_t$ (GeV)& && \omit &$\sigma_{tot}$~~{\rm (fb)}~~~~& \omit && \cr
&  && && \omit && \omit && \cr
\noalign{\hrule}
&  && && && && \cr
&  &145~~~~~&  &$610.25$~~~~~& &$582.41$~~~~~~& &$588.67$~~~~~~&  \cr
&  && && && && \cr
&  &149~~~~~&  &$276.54$~~~~~& &$250.10$~~~~~~& &$254.46$~~~~~~&  \cr
&  && && && && \cr
&  &149.5~~~~& &$195.87$~~~~~& &$178.78$~~~~~~& &$182.84$~~~~~~&  \cr
&  && && && && \cr
\noalign{\hrule}
&  && && \omit && \omit && \cr
&  && &$\sqrt s=300$ GeV& \omit && \omit &{\mathrm {Without Higgs}}& \cr
&  && && \omit && \omit && \cr
\noalign{\hrule}
\noalign{\smallskip}\cr}}$$

\centerline{\Large Table I}
\vfill
\eject
\
\vskip2.0cm
$$\vbox{\tabskip=0pt \offinterlineskip
\halign to380pt{\strut#& \vrule#\tabskip=1.em plus2.0em& \hfil#&
\vrule#& \hfil#&
\vrule#& \hfil#&
\vrule#& \hfil#&
\vrule#\tabskip=0pt\cr \noalign{\hrule}
&  && && && && \cr
&  && &$t\bar t\hskip1.2cm$& &~$t\bar t \rightarrow \bar b bW^+W^-$& &$\bar b
bW^+W^-$~~~& \cr
&  && && && && \cr
\noalign{\hrule}
&  && && \omit && \omit && \cr
&  &$m_t$ (GeV)& && \omit &$\sigma_{tot}$~~{\rm (fb)}~~~~& \omit && \cr
&  && && \omit && \omit && \cr
\noalign{\hrule}
&  && && && && \cr
&  &130~~~~~& &$654.67$~~~~~& &$665.19$~~~~~~& &$711.43$~~~~~~&  \cr
&  && && && && \cr
&  &150~~~~~& &$622.20$~~~~~& &$628.97$~~~~~~& &$663.11$~~~~~~&  \cr
&  && && && && \cr
&  &180~~~~~& &$553.60$~~~~~& &$552.51$~~~~~~& &$576.26$~~~~~~&  \cr
&  && && && && \cr
&  &200~~~~~& &$487.84$~~~~~& &$479.28$~~~~~~& &$497.63$~~~~~~&  \cr
&  && && && && \cr
\noalign{\hrule}
&  && && \omit && \omit && \cr
&  && &$\sqrt s=500$ GeV& \omit && \omit &{\mathrm {Without Higgs}}& \cr
&  && && \omit && \omit && \cr
\noalign{\hrule}
\noalign{\smallskip}\cr}}$$

\centerline{\Large Table II}
\vfill

\newpage

\setcounter{page}{0}
\vskip 2.0cm

\begin{picture}(10000,8000)
\THICKLINES
\bigphotons
\drawline\photon[\W\REG](10000,8000)[6]
\drawline\fermion[\NW\REG](\photonbackx,\photonbacky)[5000]
\drawarrow[\SE\ATBASE](\pmidx,\pmidy)
\drawline\fermion[\SW\REG](\photonbackx,\photonbacky)[5000]
\drawarrow[\SW\ATBASE](\pmidx,\pmidy)
\drawline\fermion[\NE\REG](\photonfrontx,\photonfronty)[5000]
\drawarrow[\NE\ATBASE](\pmidx,\pmidy)
\drawline\photon[\E\REG](12500,10500)[3]
\drawline\photon[\E\REG](10500,8500)[5]
\drawline\fermion[\SE\REG](10000,8000)[5000]
\drawarrow[\NW\ATBASE](\pmidx,\pmidy)
\put(-500,12000){$e^-$}
\put(-500,3000){$e^+$}
\put(14000,12000){$b$}
\put(14000,3000){$\bar b$}
\put(16000,10000){$W^+$}
\put(16000,8000){$W^-$}
\put(6500,2000){$(1)$}
\drawline\photon[\W\REG](32000,8000)[6]
\drawline\fermion[\NW\REG](\photonbackx,\photonbacky)[5000]
\drawarrow[\SE\ATBASE](\pmidx,\pmidy)
\drawline\fermion[\SW\REG](\photonbackx,\photonbacky)[5000]
\drawarrow[\SW\ATBASE](\pmidx,\pmidy)
\drawline\fermion[\NE\REG](\photonfrontx,\photonfronty)[5000]
\drawarrow[\NE\ATBASE](\pmidx,\pmidy)
\drawline\photon[\E\REG](\pmidx,\pmidy)[4]
\drawline\fermion[\SE\REG](32000,8000)[5000]
\drawarrow[\NW\ATBASE](\pmidx,\pmidy)
\drawline\photon[\E\REG](\pmidx,\pmidy)[4]
\put(21500,12000){$e^-$}
\put(21500,3000){$e^+$}
\put(36000,12000){$b$}
\put(36000,3000){$\bar b$}
\put(38250,9250){$W^+$}
\put(38250,5750){$W^-$}
\put(28500,2000){$(2)$}
\end{picture}

\vskip 2.0cm

\begin{picture}(10000,8000)
\THICKLINES
\bigphotons
\drawline\photon[\W\REG](10000,8000)[6]
\drawline\fermion[\NW\REG](\photonbackx,\photonbacky)[5000]
\drawarrow[\SE\ATBASE](\pmidx,\pmidy)
\drawline\fermion[\SW\REG](\photonbackx,\photonbacky)[5000]
\drawarrow[\SW\ATBASE](\pmidx,\pmidy)
\drawline\fermion[\NE\REG](\photonfrontx,\photonfronty)[5000]
\drawarrow[\NE\ATBASE](\pmidx,\pmidy)
\drawline\fermion[\SE\REG](10000,8000)[5000]
\drawarrow[\NW\ATBASE](\pmidx,\pmidy)
\drawline\photon[\E\REG](12500,5500)[3]
\drawline\photon[\E\REG](10500,7500)[5]
\put(-500,12000){$e^-$}
\put(-500,3000){$e^+$}
\put(14000,12000){$b$}
\put(14000,3000){$\bar b$}
\put(16000,5000){$W^-$}
\put(16000,7000){$W^+$}
\put(6500,2000){$(3)$}
\drawline\photon[\W\REG](32000,8000)[6]
\drawline\fermion[\NW\REG](\photonbackx,\photonbacky)[5000]
\drawarrow[\SE\ATBASE](\pmidx,\pmidy)
\drawline\fermion[\SW\REG](\photonbackx,\photonbacky)[5000]
\drawarrow[\SW\ATBASE](\pmidx,\pmidy)
\drawline\photon[\E\REG](23000,11000)[5]
\drawline\photon[\E\REG](25000,9000)[3]
\drawline\fermion[\NE\REG](32000,8000)[5000]
\drawarrow[\NE\ATBASE](\pmidx,\pmidy)
\drawline\fermion[\SE\REG](32000,8000)[5000]
\drawarrow[\NW\ATBASE](\pmidx,\pmidy)
\put(21500,12000){$e^-$}
\put(21500,3000){$e^+$}
\put(36000,12000){$b$}
\put(36000,3000){$\bar b$}
\put(28500,11000){$W^-$}
\put(28500,9000){$W^+$}
\put(28500,2000){$(4)$}
\end{picture}

\vskip 2.0cm

\begin{picture}(10000,8000)
\THICKLINES
\bigphotons
\drawline\photon[\W\REG](10000,8000)[6]
\drawline\fermion[\NW\REG](\photonbackx,\photonbacky)[5000]
\drawarrow[\SE\ATBASE](\pmidx,\pmidy)
\drawline\photon[\E\REG](\pmidx,\pmidy)[4]
\drawline\fermion[\SW\REG](4000,8000)[5000]
\drawarrow[\SW\ATBASE](\pmidx,\pmidy)
\drawline\photon[\E\REG](\pmidx,\pmidy)[4]
\drawline\fermion[\NE\REG](10000,8000)[5000]
\drawarrow[\NE\ATBASE](\pmidx,\pmidy)
\drawline\fermion[\SE\REG](10000,8000)[5000]
\drawarrow[\NW\ATBASE](\pmidx,\pmidy)
\put(-500,12000){$e^-$}
\put(-500,3000){$e^+$}
\put(14000,12000){$b$}
\put(14000,3000){$\bar b$}
\put(6500,9500){$W^-$}
\put(6500,5500){$W^+$}
\put(6500,2000){$(5)$}
\drawline\photon[\W\REG](32000,8000)[6]
\drawline\fermion[\NW\REG](\photonbackx,\photonbacky)[5000]
\drawarrow[\SE\ATBASE](\pmidx,\pmidy)
\drawline\photon[\E\REG](23000,5000)[5]
\drawline\photon[\E\REG](25000,7000)[3]
\drawline\fermion[\SW\REG](26000,8000)[5000]
\drawarrow[\SW\ATBASE](\pmidx,\pmidy)
\drawline\fermion[\NE\REG](32000,8000)[5000]
\drawarrow[\NE\ATBASE](\pmidx,\pmidy)
\drawline\fermion[\SE\REG](32000,8000)[5000]
\drawarrow[\NW\ATBASE](\pmidx,\pmidy)
\put(21500,12000){$e^-$}
\put(21500,3000){$e^+$}
\put(36000,12000){$b$}
\put(36000,3000){$\bar b$}
\put(28500,4500){$W^+$}
\put(28500,6500){$W^-$}
\put(28500,2000){$(6)$}
\end{picture}

\vskip 2.0cm

\begin{picture}(10000,8000)
\THICKLINES
\bigphotons
\drawline\photon[\W\REG](10000,8000)[6]
\drawline\fermion[\NW\REG](\photonbackx,\photonbacky)[5000]
\drawarrow[\SE\ATBASE](\pmidx,\pmidy)
\drawline\photon[\E\REG](\pmidx,\pmidy)[4]
\drawline\fermion[\SW\REG](4000,8000)[5000]
\drawarrow[\SW\ATBASE](\pmidx,\pmidy)
\drawline\fermion[\NE\REG](10000,8000)[5000]
\drawarrow[\NE\ATBASE](\pmidx,\pmidy)
\drawline\photon[\E\REG](\pmidx,\pmidy)[4]
\drawline\fermion[\SE\REG](10000,8000)[5000]
\drawarrow[\NW\ATBASE](\pmidx,\pmidy)
\put(-500,12000){$e^-$}
\put(-500,3000){$e^+$}
\put(14000,12000){$b$}
\put(14000,3000){$\bar b$}
\put(16000,9500){$W^+$}
\put(6500,9500){$W^-$}
\put(6500,2000){$(7)$}
\drawline\photon[\W\REG](32000,8000)[6]
\drawline\fermion[\NW\REG](\photonbackx,\photonbacky)[5000]
\drawarrow[\SE\ATBASE](\pmidx,\pmidy)
\drawline\fermion[\SW\REG](26000,8000)[5000]
\drawarrow[\SW\ATBASE](\pmidx,\pmidy)
\drawline\photon[\E\REG](\pmidx,\pmidy)[4]
\drawline\fermion[\NE\REG](32000,8000)[5000]
\drawarrow[\NE\ATBASE](\pmidx,\pmidy)
\drawline\fermion[\SE\REG](32000,8000)[5000]
\drawarrow[\NW\ATBASE](\pmidx,\pmidy)
\drawline\photon[\E\REG](\pmidx,\pmidy)[4]
\put(21500,12000){$e^-$}
\put(21500,3000){$e^+$}
\put(36000,12000){$b$}
\put(36000,3000){$\bar b$}
\put(38250,5500){$W^-$}
\put(28500,5500){$W^+$}
\put(28500,2000){$(8)$}
\end{picture}

\vskip 3.0cm
\centerline{\bf\Large Fig. 1}

\vfill
\newpage
\setcounter{page}{0}
\
\vskip 2.0cm

\begin{picture}(10000,8000)
\THICKLINES
\bigphotons
\drawline\photon[\W\REG](10000,8000)[6]
\drawline\fermion[\NW\REG](\photonbackx,\photonbacky)[5000]
\drawarrow[\SE\ATBASE](\pmidx,\pmidy)
\drawline\fermion[\SW\REG](\photonbackx,\photonbacky)[5000]
\drawarrow[\SW\ATBASE](\pmidx,\pmidy)
\drawline\fermion[\NE\REG](\photonfrontx,\photonfronty)[5000]
\drawarrow[\NE\ATBASE](\pmidx,\pmidy)
\drawvertex\photon[\E 3](\pmidx,\pmidy)[3]
\drawline\fermion[\SE\REG](10000,8000)[5000]
\drawarrow[\NW\ATBASE](\pmidx,\pmidy)
\put(-500,12000){$e^-$}
\put(-500,3000){$e^+$}
\put(14000,12000){$b$}
\put(14000,3000){$\bar b$}
\put(17300,11700){$W^+$}
\put(17300,7050){$W^-$}
\put(6500,2000){$(9)$}
\drawline\photon[\W\REG](32000,8000)[6]
\drawline\fermion[\NW\REG](\photonbackx,\photonbacky)[5000]
\drawarrow[\SE\ATBASE](\pmidx,\pmidy)
\drawline\fermion[\SW\REG](\photonbackx,\photonbacky)[5000]
\drawarrow[\SW\ATBASE](\pmidx,\pmidy)
\drawline\fermion[\NE\REG](\photonfrontx,\photonfronty)[5000]
\drawarrow[\NE\ATBASE](\pmidx,\pmidy)
\drawline\fermion[\SE\REG](32000,8000)[5000]
\drawarrow[\NW\ATBASE](\pmidx,\pmidy)
\drawvertex\photon[\E 3](\pmidx,\pmidy)[3]
\put(21500,12000){$e^-$}
\put(21500,3000){$e^+$}
\put(36000,12000){$b$}
\put(36000,3000){$\bar b$}
\put(39300,8200){$W^+$}
\put(39300,3550){$W^-$}
\put(28000,2000){$(10)$}
\end{picture}

\vskip 2.0cm

\begin{picture}(10000,8000)
\THICKLINES
\bigphotons
\drawline\photon[\W\REG](11000,0)[4]
\drawline\fermion[\N\REG](\photonbackx,\photonbacky)[8000]
\drawarrow[\S\ATBASE](\pmidx,\pmidy)
\drawline\fermion[\NW\REG](\fermionbackx,\fermionbacky)[5000]
\drawarrow[\SE\ATBASE](\pmidx,\pmidy)
\drawline\fermion[\SW\REG](7000,0)[5000]
\drawarrow[\SW\ATBASE](\pmidx,\pmidy)
\drawline\fermion[\NE\REG](11000,0)[4000]
\drawarrow[\NE\ATBASE](\pmidx,\pmidy)
\drawline\fermion[\SE\REG](\fermionfrontx,\fermionfronty)[4000]
\drawarrow[\NW\ATBASE](\pmidx,\pmidy)
\drawvertex\photon[\E 3](7000,8000)[4]
\put(2500,12000){$e^-$}
\put(2500,-5000){$e^+$}
\put(14150,10500){$W^+$}
\put(14150,4750){$W^-$}
\put(14300,2750){$b$}
\put(14300,-3750){$\bar b$}
\put(8000,-7000){$(11)$}
\drawline\photon[\W\REG](33000,8000)[4]
\drawline\fermion[\NW\REG](\photonbackx,\photonbacky)[5000]
\drawarrow[\SE\ATBASE](\pmidx,\pmidy)
\drawline\fermion[\S\REG](\photonbackx,\photonbacky)[8000]
\drawarrow[\S\ATBASE](\pmidx,\pmidy)
\drawline\fermion[\SW\REG](\fermionbackx,\fermionbacky)[5000]
\drawarrow[\SW\ATBASE](\pmidx,\pmidy)
\drawline\fermion[\NE\REG](\photonfrontx,\photonfronty)[4000]
\drawarrow[\NE\ATBASE](\pmidx,\pmidy)
\drawline\fermion[\SE\REG](33000,8000)[4000]
\drawarrow[\NW\ATBASE](\pmidx,\pmidy)
\drawvertex\photon[\E 3](29000,0)[4]
\put(24500,12000){$e^-$}
\put(24500,-5000){$e^+$}
\put(36300,10750){$b$}
\put(36300,4250){$\bar b$}
\put(36150,2500){$W^+$}
\put(36150,-3250){$W^-$}
\put(30000,-7000){$(12)$}
\end{picture}

\vskip 5.0cm
\centerline{\bf\Large Fig. 1 (Continued)}

\vfill
\newpage
\setcounter{page}{0}
\
\vskip 2.0cm

\begin{picture}(10000,8000)
\THICKLINES
\bigphotons
\drawline\photon[\W\REG](10000,8000)[6]
\drawline\fermion[\NW\REG](\photonbackx,\photonbacky)[5000]
\drawarrow[\SE\ATBASE](\pmidx,\pmidy)
\drawline\fermion[\SW\REG](\photonbackx,\photonbacky)[5000]
\drawarrow[\SW\ATBASE](\pmidx,\pmidy)
\drawline\photon[\NE\REG](\photonfrontx,\photonfronty)[6]
\drawline\photon[\E\REG](12300,10000)[3]
\drawline\fermion[\NE\REG](\photonbackx,\photonbacky)[3000]
\drawarrow[\NE\ATBASE](\pmidx,\pmidy)
\drawline\fermion[\SE\REG](\photonbackx,\photonbacky)[3000]
\drawarrow[\NW\ATBASE](\pmidx,\pmidy)
\drawline\photon[\SE\REG](10000,8000)[6]
\put(-500,12000){$e^-$}
\put(-500,3000){$e^+$}
\put(14000,12000){$W^+$}
\put(14000,3000){$W^-$}
\put(17650,12000){$b$}
\put(17650,6750){$\bar b$}
\put(6000,2000){$(13)$}
\drawline\photon[\W\REG](32000,8000)[6]
\drawline\fermion[\NW\REG](\photonbackx,\photonbacky)[5000]
\drawarrow[\SE\ATBASE](\pmidx,\pmidy)
\drawline\fermion[\SW\REG](\photonbackx,\photonbacky)[5000]
\drawarrow[\SW\ATBASE](\pmidx,\pmidy)
\drawline\photon[\NE\REG](\photonfrontx,\photonfronty)[6]
\drawline\photon[\SE\REG](32000,8000)[6]
\drawline\photon[\E\REG](\pmidx,\pmidy)[3]
\drawline\fermion[\NE\REG](\photonbackx,\photonbacky)[3000]
\drawarrow[\NE\ATBASE](\pmidx,\pmidy)
\drawline\fermion[\SE\REG](\photonbackx,\photonbacky)[3000]
\drawarrow[\NW\ATBASE](\pmidx,\pmidy)
\put(21500,12000){$e^-$}
\put(21500,3000){$e^+$}
\put(36000,12000){$W^+$}
\put(36000,3000){$W^-$}
\put(39300,8400){$b$}
\put(39300,2900){$\bar b$}
\put(28000,2000){$(14)$}
\end{picture}

\vskip 2.0cm

\begin{picture}(10000,8000)
\THICKLINES
\bigphotons
\drawline\photon[\W\REG](21000,8000)[5]
\drawline\fermion[\NW\REG](\photonbackx,\photonbacky)[5000]
\drawarrow[\SE\ATBASE](\pmidx,\pmidy)
\drawline\fermion[\SW\REG](\photonbackx,\photonbacky)[5000]
\drawarrow[\SW\ATBASE](\pmidx,\pmidy)
\drawline\photon[\NE\REG](\photonfrontx,\photonfronty)[6]
\drawline\photon[\E\REG](\photonfrontx,\photonfronty)[5]
\drawline\fermion[\NE\REG](\photonbackx,\photonbacky)[3000]
\drawarrow[\NE\ATBASE](\pmidx,\pmidy)
\drawline\fermion[\SE\REG](\photonbackx,\photonbacky)[3000]
\drawarrow[\NW\ATBASE](\pmidx,\pmidy)
\drawline\photon[\SE\REG](21000,8000)[6]
\put(11500,12000){$e^-$}
\put(11500,3000){$e^+$}
\put(25000,12000){$W^+$}
\put(25000,3000){$W^-$}
\put(28400,10250){$b$}
\put(28400,4750){$\bar b$}
\put(20000,2000){$(15)$}
\end{picture}

\vskip 3.0cm
\centerline{\bf\Large Fig. 1 (Continued)}

\vfill
\newpage
\setcounter{page}{0}
\
\vskip 2.0cm

\begin{picture}(10000,8000)
\THICKLINES
\bigphotons
\drawline\photon[\W\REG](9000,8000)[5]
\drawline\fermion[\NW\REG](\photonbackx,\photonbacky)[5000]
\drawarrow[\SE\ATBASE](\pmidx,\pmidy)
\drawline\fermion[\SW\REG](\photonbackx,\photonbacky)[5000]
\drawarrow[\SW\ATBASE](\pmidx,\pmidy)
\drawline\photon[\NE\REG](\photonfrontx,\photonfronty)[6]
\drawline\photon[\SE\REG](9000,8000)[4]
\drawline\fermion[\NE\REG](\photonbackx,\photonbacky)[3000]
\drawarrow[\NE\ATBASE](\pmidx,\pmidy)
\drawline\fermion[\SE\REG](\photonbackx,\photonbacky)[3000]
\drawarrow[\NW\ATBASE](\pmidx,\pmidy)
\drawline\photon[\E\REG](\pmidx,\pmidy)[3]
\put(-500,12000){$e^-$}
\put(-500,3000){$e^+$}
\put(13000,12000){$W^+$}
\put(15900,4200){$W^-$}
\put(13750,8000){$b$}
\put(13750,2250){$\bar b$}
\put(6000,2000){$(16)$}
\drawline\photon[\W\REG](31000,8000)[5]
\drawline\fermion[\NW\REG](\photonbackx,\photonbacky)[5000]
\drawarrow[\SE\ATBASE](\pmidx,\pmidy)
\drawline\fermion[\SW\REG](\photonbackx,\photonbacky)[5000]
\drawarrow[\SW\ATBASE](\pmidx,\pmidy)
\drawline\photon[\NE\REG](\photonfrontx,\photonfronty)[4]
\drawline\fermion[\NE\REG](\photonbackx,\photonbacky)[3000]
\drawarrow[\NE\ATBASE](\pmidx,\pmidy)
\drawline\photon[\E\REG](\pmidx,\pmidy)[3]
\drawline\fermion[\SE\REG](\fermionfrontx,\fermionfronty)[3000]
\drawarrow[\NW\ATBASE](\pmidx,\pmidy)
\drawline\photon[\SE\REG](31000,8000)[6]
\put(21500,12000){$e^-$}
\put(21500,3000){$e^+$}
\put(37800,11250){$W^+$}
\put(35000,3000){$W^-$}
\put(35750,12750){$b$}
\put(35750,7000){$\bar b$}
\put(28000,2000){$(17)$}
\end{picture}

\vskip 2.0cm

\begin{picture}(10000,8000)
\THICKLINES
\bigphotons
\drawline\photon[\W\REG](9000,8000)[5]
\drawline\fermion[\NW\REG](\photonbackx,\photonbacky)[5000]
\drawarrow[\SE\ATBASE](\pmidx,\pmidy)
\drawline\photon[\E\REG](\pmidx,\pmidy)[3]
\drawline\fermion[\SW\REG](\fermionfrontx,\fermionfronty)[5000]
\drawarrow[\SW\ATBASE](\pmidx,\pmidy)
\drawline\photon[\NE\REG](9000,8000)[6]
\drawline\photon[\SE\REG](9000,8000)[4]
\drawline\fermion[\NE\REG](\photonbackx,\photonbacky)[3000]
\drawarrow[\NE\ATBASE](\pmidx,\pmidy)
\drawline\fermion[\SE\REG](\photonbackx,\photonbacky)[3000]
\drawarrow[\NW\ATBASE](\pmidx,\pmidy)
\put(-500,12000){$e^-$}
\put(-500,3000){$e^+$}
\put(13000,12000){$W^+$}
\put(5500,9500){$W^-$}
\put(13750,8000){$b$}
\put(13750,2250){$\bar b$}
\put(6000,2000){$(18)$}
\drawline\photon[\W\REG](31000,8000)[5]
\drawline\fermion[\NW\REG](\photonbackx,\photonbacky)[5000]
\drawarrow[\SE\ATBASE](\pmidx,\pmidy)
\drawline\fermion[\SW\REG](\fermionfrontx,\fermionfronty)[5000]
\drawarrow[\SW\ATBASE](\pmidx,\pmidy)
\drawline\photon[\E\REG](\pmidx,\pmidy)[3]
\drawline\photon[\SE\REG](31000,8000)[6]
\drawline\photon[\NE\REG](31000,8000)[4]
\drawline\fermion[\NE\REG](\photonbackx,\photonbacky)[3000]
\drawarrow[\NE\ATBASE](\pmidx,\pmidy)
\drawline\fermion[\SE\REG](\photonbackx,\photonbacky)[3000]
\drawarrow[\NW\ATBASE](\pmidx,\pmidy)
\put(21500,12000){$e^-$}
\put(21500,3000){$e^+$}
\put(35000,3000){$W^-$}
\put(27500,5750){$W^+$}
\put(35750,12500){$b$}
\put(35750,7000){$\bar b$}
\put(28000,2000){$(19)$}
\end{picture}

\vskip 3.0cm
\centerline{\bf\Large Fig. 1 (Continued)}

\vfill
\newpage
\setcounter{page}{0}
\
\vskip 2.0cm

\begin{picture}(10000,8000)
\THICKLINES
\bigphotons
\drawline\photon[\W\REG](10000,8000)[6]
\drawline\fermion[\NW\REG](\photonbackx,\photonbacky)[5000]
\drawarrow[\SE\ATBASE](\pmidx,\pmidy)
\drawline\fermion[\SW\REG](\photonbackx,\photonbacky)[5000]
\drawarrow[\SW\ATBASE](\pmidx,\pmidy)
\drawline\fermion[\NE\REG](\photonfrontx,\photonfronty)[5000]
\drawarrow[\NE\ATBASE](\pmidx,\pmidy)
\seglength=1416  \gaplength=300  
\drawline\scalar[\E\REG](\pmidx,\pmidy)[2]
\drawline\photon[\NE\REG](\scalarbackx,\scalarbacky)[3]
\drawline\photon[\SE\REG](\scalarbackx,\scalarbacky)[3]
\drawline\fermion[\SE\REG](10000,8000)[5000]
\drawarrow[\NW\ATBASE](\pmidx,\pmidy)
\put(-500,12000){$e^-$}
\put(-500,3000){$e^+$}
\put(14000,12000){$b$}
\put(14000,3000){$\bar b$}
\put(17250,11250){$W^+$}
\put(17250,7500){$W^-$}
\put(6000,2000){$(20)$}
\drawline\photon[\W\REG](32000,8000)[6]
\drawline\fermion[\NW\REG](\photonbackx,\photonbacky)[5000]
\drawarrow[\SE\ATBASE](\pmidx,\pmidy)
\drawline\fermion[\SW\REG](\photonbackx,\photonbacky)[5000]
\drawarrow[\SW\ATBASE](\pmidx,\pmidy)
\drawline\fermion[\NE\REG](\photonfrontx,\photonfronty)[5000]
\drawarrow[\NE\ATBASE](\pmidx,\pmidy)
\drawline\fermion[\SE\REG](32000,8000)[5000]
\drawarrow[\NW\ATBASE](\pmidx,\pmidy)
\seglength=1416  \gaplength=300  
\drawline\scalar[\E\REG](\pmidx,\pmidy)[2]
\drawline\photon[\NE\REG](\scalarbackx,\scalarbacky)[3]
\drawline\photon[\SE\REG](\scalarbackx,\scalarbacky)[3]
\put(21500,12000){$e^-$}
\put(21500,3000){$e^+$}
\put(36000,12000){$b$}
\put(36000,3000){$\bar b$}
\put(39250,7750){$W^+$}
\put(39250,4000){$W^-$}
\put(28000,2000){$(21)$}
\end{picture}

\vskip 2.0cm

\begin{picture}(10000,8000)
\THICKLINES
\bigphotons
\drawline\photon[\W\REG](10000,8000)[6]
\drawline\fermion[\NW\REG](\photonbackx,\photonbacky)[5000]
\drawarrow[\SE\ATBASE](\pmidx,\pmidy)
\drawline\fermion[\SW\REG](\photonbackx,\photonbacky)[5000]
\drawarrow[\SW\ATBASE](\pmidx,\pmidy)
\drawline\photon[\NE\REG](\photonfrontx,\photonfronty)[6]
\seglength=1416  \gaplength=300  
\drawline\scalar[\E\REG](\pmidx,\pmidy)[2]
\drawline\fermion[\NE\REG](\scalarbackx,\scalarbacky)[3000]
\drawarrow[\NE\ATBASE](\pmidx,\pmidy)
\drawline\fermion[\SE\REG](\scalarbackx,\scalarbacky)[3000]
\drawarrow[\NW\ATBASE](\pmidx,\pmidy)
\drawline\photon[\SE\REG](10000,8000)[6]
\put(-500,12000){$e^-$}
\put(-500,3000){$e^+$}
\put(14000,12000){$W^+$}
\put(14000,3000){$W^-$}
\put(17500,12250){$b$}
\put(17500,6750){$\bar b$}
\put(6000,2000){$(22)$}
\drawline\photon[\W\REG](32000,8000)[6]
\drawline\fermion[\NW\REG](\photonbackx,\photonbacky)[5000]
\drawarrow[\SE\ATBASE](\pmidx,\pmidy)
\drawline\fermion[\SW\REG](\photonbackx,\photonbacky)[5000]
\drawarrow[\SW\ATBASE](\pmidx,\pmidy)
\drawline\photon[\NE\REG](\photonfrontx,\photonfronty)[6]
\drawline\photon[\SE\REG](32000,8000)[6]
\seglength=1416  \gaplength=300  
\drawline\scalar[\E\REG](\pmidx,\pmidy)[2]
\drawline\fermion[\NE\REG](\scalarbackx,\scalarbacky)[3000]
\drawarrow[\NE\ATBASE](\pmidx,\pmidy)
\drawline\fermion[\SE\REG](\scalarbackx,\scalarbacky)[3000]
\drawarrow[\NW\ATBASE](\pmidx,\pmidy)
\put(21500,12000){$e^-$}
\put(21500,3000){$e^+$}
\put(36000,12000){$W^+$}
\put(36000,3000){$W^-$}
\put(39500,8500){$b$}
\put(39500,3000){$\bar b$}
\put(28000,2000){$(23)$}
\end{picture}

\vskip 2.0cm

\begin{picture}(10000,8000)
\THICKLINES
\bigphotons
\drawline\photon[\W\REG](9000,8000)[5]
\drawline\fermion[\NW\REG](\photonbackx,\photonbacky)[5000]
\drawarrow[\SE\ATBASE](\pmidx,\pmidy)
\drawline\photon[\E\REG](\pmidx,\pmidy)[3]
\drawline\fermion[\SW\REG](\fermionfrontx,\fermionfronty)[5000]
\drawarrow[\SW\ATBASE](\pmidx,\pmidy)
\drawline\photon[\NE\REG](9000,8000)[6]
\seglength=1416  \gaplength=300  
\drawline\scalar[\SE\REG](\photonfrontx,\photonfronty)[2]
\drawline\fermion[\NE\REG](\scalarbackx,\scalarbacky)[3000]
\drawarrow[\NE\ATBASE](\pmidx,\pmidy)
\drawline\fermion[\SE\REG](\scalarbackx,\scalarbacky)[3000]
\drawarrow[\NW\ATBASE](\pmidx,\pmidy)
\put(-500,12000){$e^-$}
\put(-500,3000){$e^+$}
\put(13000,12000){$W^+$}
\put(5650,9500){$W^-$}
\put(13850,8000){$b$}
\put(13850,2750){$\bar b$}
\put(6000,2000){$(24)$}
\drawline\photon[\W\REG](31000,8000)[5]
\drawline\fermion[\NW\REG](\photonbackx,\photonbacky)[5000]
\drawarrow[\SE\ATBASE](\pmidx,\pmidy)
\drawline\fermion[\SW\REG](\fermionfrontx,\fermionfronty)[5000]
\drawarrow[\SW\ATBASE](\pmidx,\pmidy)
\drawline\photon[\E\REG](\pmidx,\pmidy)[3]
\drawline\photon[\SE\REG](31000,8000)[6]
\seglength=1416  \gaplength=300  
\drawline\scalar[\NE\REG](\photonfrontx,\photonfronty)[2]
\drawline\fermion[\NE\REG](\scalarbackx,\scalarbacky)[3000]
\drawarrow[\NE\ATBASE](\pmidx,\pmidy)
\drawline\fermion[\SE\REG](\scalarbackx,\scalarbacky)[3000]
\drawarrow[\NW\ATBASE](\pmidx,\pmidy)
\put(21500,12000){$e^-$}
\put(21500,3000){$e^+$}
\put(35000,3000){$W^-$}
\put(27650,5900){$W^+$}
\put(35750,12250){$b$}
\put(35750,6750){$\bar b$}
\put(28000,2000){$(25)$}
\end{picture}

\vskip 2.0cm

\begin{picture}(10000,8000)
\THICKLINES
\bigphotons
\drawline\photon[\W\REG](9000,8000)[5]
\drawline\fermion[\NW\REG](\photonbackx,\photonbacky)[5000]
\drawarrow[\SE\ATBASE](\pmidx,\pmidy)
\drawline\fermion[\SW\REG](\photonbackx,\photonbacky)[5000]
\drawarrow[\SW\ATBASE](\pmidx,\pmidy)
\seglength=1416  \gaplength=300  
\drawline\scalar[\NE\REG](\photonfrontx,\photonfronty)[3]
\drawline\photon[\NE\REG](\scalarbackx,\scalarbacky)[4]
\drawline\photon[\SE\REG](\scalarbackx,\scalarbacky)[4]
\drawline\photon[\SE\REG](9000,8000)[4]
\drawline\fermion[\NE\REG](\photonbackx,\photonbacky)[3000]
\drawarrow[\NE\ATBASE](\pmidx,\pmidy)
\drawline\fermion[\SE\REG](\photonbackx,\photonbacky)[3000]
\drawarrow[\NW\ATBASE](\pmidx,\pmidy)
\put(-500,12000){$e^-$}
\put(-500,3000){$e^+$}
\put(15250,13250){$W^+$}
\put(15250,8500){$W^-$}
\put(13750,8000){$b$}
\put(13750,2250){$\bar b$}
\put(6000,2000){$(26)$}
\drawline\photon[\W\REG](31000,8000)[5]
\drawline\fermion[\NW\REG](\photonbackx,\photonbacky)[5000]
\drawarrow[\SE\ATBASE](\pmidx,\pmidy)
\drawline\fermion[\SW\REG](\photonbackx,\photonbacky)[5000]
\drawarrow[\SW\ATBASE](\pmidx,\pmidy)
\seglength=1416  \gaplength=300  
\drawline\scalar[\NE\REG](\photonfrontx,\photonfronty)[3]
\drawline\fermion[\NE\REG](\scalarbackx,\scalarbacky)[3000]
\drawarrow[\NE\ATBASE](\pmidx,\pmidy)
\drawline\fermion[\SE\REG](\scalarbackx,\scalarbacky)[3000]
\drawarrow[\NW\ATBASE](\pmidx,\pmidy)
\drawline\photon[\SE\REG](31000,8000)[4]
\drawline\photon[\NE\REG](\photonbackx,\photonbacky)[4]
\drawline\photon[\SE\REG](\photonfrontx,\photonfronty)[4]
\put(21500,12000){$e^-$}
\put(21500,3000){$e^+$}
\put(37000,13500){$b$}
\put(37000,8750){$\bar b$}
\put(36250,7350){$W^+$}
\put(36250,2250){$W^-$}
\put(28000,2000){$(27)$}
\end{picture}

\vskip 3.0cm
\centerline{\bf\Large Fig. 1 (Continued)}


\begin{thebibliography}{1}

\bibitem{ee500} Proc. of the Workshop ``{\it $e^+e^-$ Collisions at
500 GeV. The Physics Potential}\ '',
Munich, Annecy, Hamburg, 3--4 February 1991, ed. P.M.~Zerwas, DESY pub.
92--123A/B,
Aug. 1992.

\bibitem{JLC} Proc. of the I Workshop on Japan Linear Collider (JLC), KEK
1989,
KEK-report 90--2;\\
Proc. of the II Workshop on Japan Linear Collider (JLC), KEK  1990,
KEK-report 91--10.

\bibitem{widthtop} I.~Bigi, Y.~Dokshitzer, V.A.~Khoze, J.~Kuhn and P.~Zerwas,
{\it Phys. Lett.} {\bf B181} (1986) 157;\\
R.~Kleiss and W.J.~Stirling,
{\it Z. Phys.} {\bf C40} (1988) 419;\\
M.~Je\.zabek and J.H.~K\"uhn,
{\it preprint} TTP93--4, Feb. 1993.

\bibitem{qcdcorr}  J.~Jersak, E.~Laermann and P.M.~Zerwas,
{\it Phys. Rev.} {\bf D25} (1982) 263.

\bibitem{ewcorr} A.~Denner and T.~Sack,
{\it Nucl. Phys.} {\bf B358} (1991) 46;\\
W.~Beenakker, S.C. van der Marck and W.~Hollik,
{\it Nucl. Phys.} {\bf B365} (1991) 24.

\bibitem{ee500exp} G. Bagliesi {\it et al.} in ref.\cite{ee500}.

\bibitem{ttthreshold1}
V.S.~Fadin and O.I.~Yakovlev, {\it Sov. J. Nucl. Phys.} {\bf 23} (1991) 1117.

\bibitem{ttthreshold2}
J.M.~Strassler and M.E.~Peskin,
{\it Phys. Rev.} {\bf D43} (1991) 1500.

\bibitem{ttthreshold3}
M.~Je\.zabek, J.H.~K\"uhn and T.~Teubner,
{\it Z. Phys.} {\bf C56} (1992) 653.

\bibitem{ttthreshold4}
Y.~Sumino, K.~Fujii, K.~Hagiwara, H.~Murayama and C.--K.~Ng
{\it Tokyo Univ. preprint} UT--594.

\bibitem{ttthreshold5}
M.~Je\.zabek and T.~Teubner,
{\it Karlsruhe Univ. preprint} TTP 92--38, Dec. 1992.

\bibitem{stiegler} U.~Stiegler, {\it Trieste preprint}
INFN/AE-91/09, (1991).

\bibitem{singletop} S.~Ambrosanio and B.~Mele, {\it preprint
n.979 Univ. Roma ``La Sapienza''}, Nov. 1993;\\
M.~Raidal and R.~Vuopionper\" a, {\it Phys. Lett.} {\bf B318} (1993) 237;\\
O.~Panella, G.~Pancheri and Y.N.~Srivastava,
{\it Phys. Lett.} {\bf B318} (1993) 241.

\bibitem{wisc} V.~Barger, K.~Chung, B.A.~Knielh and
R.J.N.~Philips,
{\it Phys. Rev.} {\bf D46} (1992) 3725.

\bibitem{method} A.~Ballestrero, E.~Maina,
{\it Turin Univ. preprint} DFTT 76/93, Dec. 1993.

\bibitem{hz} K.~Hagiwara and D.~Zeppenfeld,
{\it Nucl. Phys.} {\bf B274} (1986) 1.

\bibitem{ks}
F.A.~Berends, P.H. Daverveldt and R.~Kleiss,
{\it Nucl. Phys.} {\bf B253} (1985) 441;\\
R.~Kleiss and W.J.~Stirling,
{\it Nucl. Phys.} {\bf B262} (1985) 235;\\
C.~Mana and M.~Martinez,
{\it Nucl. Phys.} {\bf B287} (1987) 601.

\bibitem{W_width}
M.~Je\.zabek and  J.H.~K\"uhn,
{\it Nucl. Phys.} {\bf B314} (1989) 1;\\
C.~Gao, C.~Lu and W.~Lu,
{\it U. of Minnesota preprint} TPI-MINN-92/13-T, Mar. 1992.


\bibitem{ee500WWZ} J. Kalinowski in ref.\cite{ee500}.

\end{thebibliography}
\end{document}